\documentclass[a4paper,useAMS,usenatbib]{mnras}

\usepackage{graphics,epsfig,psfig}
\usepackage{todonotes}
\usepackage[normalem]{ulem}
\usepackage{xcolor}
\usepackage{float}
\usepackage{dcolumn}
\usepackage{kantlipsum,widetext}
\usepackage[style=base, singlelinecheck=off, font=small, labelfont=bf, 
justification=raggedright]{caption}
\usepackage{subfigure}
\usepackage[]{inputenc,amssymb}
\usepackage{natbib}
\usepackage{url}
\usepackage{orcidlink}
\usepackage{widetext}
\usepackage{amsmath}
\usepackage{cancel}

  \makeatletter
\def\@fnsymbol#1{\ensuremath{\ifcase#1\or \dagger\or \ddagger\or
   \mathsection\or \mathparagraph\or \|\or **\or \dagger\dagger
   \or \ddagger\ddagger \else\@ctrerr\fi}}
    \makeatother
\def \be{\begin{equation}}
\def \ee{\end{equation}}
\def \bea{\begin{eqnarray}}
\def \eea{\end{eqnarray}}

\def\apj{ApJ}

\def\aap{A\&A}
\definecolor{webgreen}{rgb}{0,.5,0}
\definecolor{webbrown}{rgb}{.6,0,0}

\def\aap{A\&A}
\def\apj{ApJ}
\def\apjs{ApJS}
\def\apjl{ApJL}

\def\pasa{Publications of the Astronomical Society of Australia}

\title[SGWB and its non-stationary properties]{Non-Stationary Astrophysical Stochastic Gravitational-Wave Background: A New Probe to the High Redshift Population of Binary Black Holes}

\author[Sah and Mukherjee]{Mohit Raj Sah$^{1}$\thanks{mohit.sah@tifr.res.in}\orcidlink{0009-0005-9881-1788},
Suvodip Mukherjee$^{1}$\thanks{suvodip.mukherjee@tifr.res.in}\orcidlink{0000-0002-3373-5236}\\
$^{1}$Department of Astronomy and Astrophysics, Tata Institute of Fundamental Research, Mumbai 400005, India}

\pubyear{2023}

\begin{document}
\label{firstpage}
\pagerange{\pageref{firstpage}--\pageref{lastpage}}
\maketitle

\label{firstpage}

\begin{abstract}

The astrophysical Stochastic Gravitational Wave Background (SGWB) originates from the mergers of compact binary objects that are otherwise undetected as individual events, along with other sources such as supernovae, magnetars, etc. The individual GW signal is time-varying over a time scale that depends on the chirp mass of the coalescing binaries. Another timescale that plays a role is the timescale at which the sources repeat, which depends on the merger rate. The combined effect of these two leads to a breakdown of the time-translation symmetry of the observed SGWB and a correlation between different frequency modes in the signal covariance matrix of the SGWB. Using an ensemble of SGWB due to binary black hole coalescence, calculated using simulations of different black hole mass distributions and merger rates, we show how the structure of the signal covariance matrix varies. This structure in the signal covariance matrix brings additional information about the sources on top of the power spectrum. We show that there is a significant improvement in the Figure of Merit by using this additional information in comparison to only power spectrum estimation for the LIGO-Virgo-KAGRA (LVK) network of detectors with the design sensitivity noise with two years of observation. The inclusion of the off-diagonal correlation in the covariance of the SGWB in the data analysis pipelines will be beneficial in the quest for the SGWB signal in LVK frequency bands as well as in lower frequencies and in getting an insight into its origin.

\end{abstract}


\section{Introduction}\label{intro}

Gravitational waves (GW) are a novel cosmic messenger that can help us answer some of the most important questions in astrophysics, cosmology, and fundamental physics \citep{thorne1995gravitational,sathyaprakash2009physics,bailes2021gravitational,perkins2021probing, Berti:2022wzk, Abdalla:2022yfr, LISACosmologyWorkingGroup:2022jok,mastrogiovanni2022cosmology, Adhikari:2022sve}. The first direct detection of gravitational waves (GW150914) was made by the LIGO-Virgo Collaboration \citep{abbott2016observation} in September 2015 which came from the merger of two black holes (BHs). So far, the LVK has detected close to 90 compact binary mergers, which include binary black holes (BBHs), binary neutron stars (BNSs), and neutron star-black holes (NSBHs) mergers \citep{abbott2021population}. Most of these sources are located below the redshift of $z=1$ for fiducial Planck cosmology \citep{abbott2021population}. The present detectors can only detect sources individually that are located below the redshift of one with a matched filtering SNR above 8 \citep{abbott2019gwtc}. All the unresolved events appear as the stochastic gravitational wave background (SGWB) signal 
\citep{apreda2001supersymmetric,zhu2011stochastic,romano2017detection}. The SGWB can be detected by ground-based detectors using a technique called cross-correlation which is the conventional approach to detect the SGWB signal \citep{allen1999detecting,thrane2013sensitivity,thrane2009probing}. Other search techniques include The Bayesian Search (TBS) \citep{smith2018optimal} and Cross-Correlation Intermediate (CCI) search \citep{coyne2016cross}. 
The Pulsar Timing Array (PTA) search \citep{burke2019astrophysics, verbiest2022pulsar, manchester2013international}, operating in the nano-Hertz range, relies on the correlated signatures in the pulse arrival times of a set of pulsars. Recently, the PTA collaborations 
 have reported the detection of the nano-Hertz gravitational waves \citep{zic2023parkes, agazie2023nanograv, antoniadis2023second, lee2023searching}. However, in the LVK, we have only been able to place an upper limit on the SGWB \citep{abbott2021upper,abbott2021search}. Nevertheless, we expect to detect the SGWB signal with future upgraded detectors \citep{thrane2013sensitivity,christensen2018stochastic,renzini2022stochastic,mentasti2023intrinsic,suresh2021jointly}. The astrophysical SGWB  is a combination of various types of sources such as compact binary mergers, supernovae, magnetars, etc \citep{buonanno2005stochastic,chowdhury2021stochastic}. However, in this paper, we concentrate solely on examining the SGWB resulting from the coalescence of BBH detected by ground-based detectors. This work can also be extended to other types of sources as well as detectors. 

The SGWB is a probe to the high redshift universe using which we can study the properties of compact objects and explore their formation channels \citep{mukherjee2021can,bavera2022stochastic,lehoucq2023astrophysical,babak2023stochastic}. One advantage of SGWB over individual source detection is that SGWB can explore deep into high redshift for sources over a large range of compact object masses ranging from sub-solar mass to super-solar mass, whose origin can be primordial or astrophysical. The merger rate of compact objects of astrophysical origin, as well as their mass and spin distributions, depends on the stellar properties of galaxies (star formation rate, stellar mass, stellar metallicity, etc.) and also the formation channels \citep{belczynski2002comprehensive,renzo2020sensitivity,bethe1998evolution,mukherjee2022redshift,spera2019merging,dorozsmai2022importance,kruckow2018progenitors,srinivasan2023understanding}. As a result, the SGWB signal and its connection with the stellar properties of the galaxies can provide rich information about the high redshift Universe in a way complementary to the electromagnetic probes. SGWB can also help us distinguish different formation channels
of BHs. It is a powerful probe to distinguish astrophysical black holes (ABHs) from primordial black holes (PBHs) \citep{mukherjee2021can}. The merger rate of ABHs is expected to follow the star formation history.
In contrast, PBHs, being formed in the very early universe, are expected to exhibit significantly different merger rates. The merger rate of PBHs is going to dominate over ABHs at high redshifts. This will appear as a distinguishable signature on the SGWB spectrum \citep{mukherjee2021can, Mukherjee:2021itf, atal2022constraining}.

The number of mergers contributing to the SGWB in a given time interval is expected to follow the Poisson distribution \citep{mandel2010compact,bulik2011ic10,dvorkin2018exploring}. This leads to fluctuations in the number of events with time \citep{mukherjee2020time, Mukherjee:2020jxa, Braglia:2022icu, ginat2023frequency, ginat2020probability}. The amount of fluctuation will depend on the merger rate and observation time. Along with that, the mass distribution of the compact objects contributing to the background can also contribute to the temporal fluctuations in the SGWB. Recently, a new technique has been proposed to search for this signal from the background using spectrogram analysis \citep{Dey:2023oui}. Furthermore, there are other proposed techniques aimed at distinguishing these signals based on their statistical properties \citep{smith2018optimal,buscicchio2023detecting,PhysRevD.107.103026}.

The measurement of the SGWB signal is typically performed through the cross-correlation of the GW strain between two detectors. Such an estimator is optimal for types of signal which is continuous and Gaussian. However, the SGWB signal of astrophysical origin is expected to exhibit temporal dependence and break stationarity due to two effects: (i) the time dependence of the individual signal, which relies on the masses of the GW sources, and (ii) the time scale over which the signal repeats.  Despite the temporal variations, the SGWB signal will appear statistically time-translation symmetric over a large timescale, as the astrophysical source population properties contributing to the signal will not change over the time scale over which observations are made.

\begin{figure*}
    \centering
    \includegraphics[width=\textwidth]{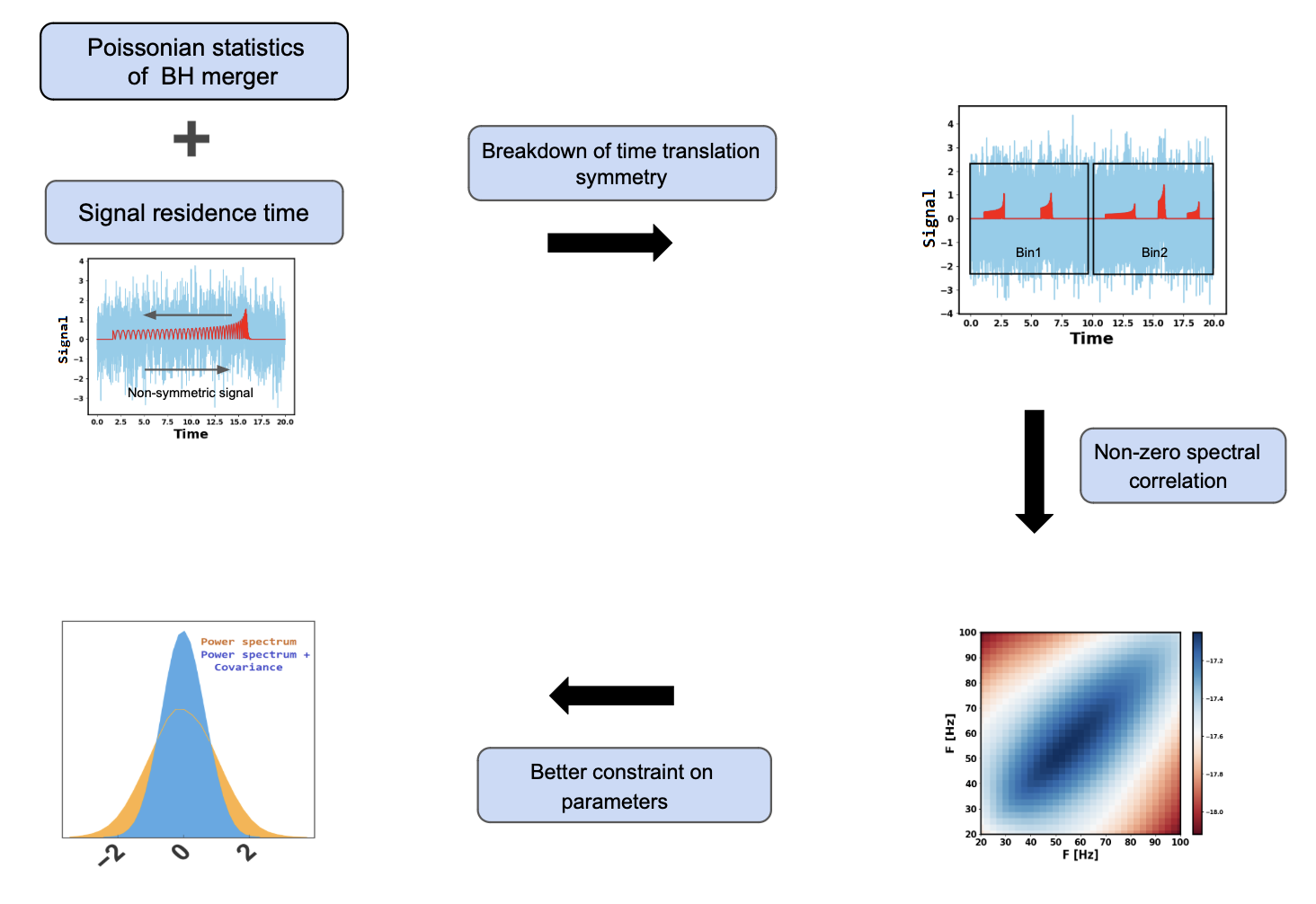}
  
    \caption{A schematic diagram explaining the concept behind this work. The chirping behaviour of the GW sources and its merger rate breaks the time-translation symmetry in the data. This leads to a correlation between different frequency modes in the covariance matrix in the Fourier space. The inclusion of this spectral correlation between different frequencies can improve our constraints on the parameters of the sources contributing to the background (The axes are in arbitrary units).}
    \label{Mot1}
\end{figure*}

The presence of a non-stationary SGWB signal leads to a correlation between two different frequency modes of the signal, and as a result, the signal covariance matrix no longer remains diagonal but contains non-zero off-diagonal terms. In Fig. \ref{Mot1} we illustrate how the non-stationary SGWB signal can give rise to correlations between different frequencies in the covariance matrix and how this spectral covariance can enhance our estimates of the parameters. The Poissonian nature of the BH merger and the chirp signal breaks the time translation symmetry of the signal. However, the noise remains stationary and therefore retains time translational symmetry\footnote{We have discussed later about the non-stationary noise and correlation between frequencies for a signal covariance matrix and noise covariance matrix.}. As a result, the spectral covariance matrix of the signal contains non-zero diagonal elements, while the noise spectral covariance matrix remains diagonal. Therefore, the off-diagonal term in the spectral covariance matrix remains uncontaminated from noise. This allows the spectral covariance to provide additional information to put a better constraint on the GW source parameters.

Using a simulation-based approach to estimate SGWB for different populations, developed in this work, we estimate the SGWB signal and show that there is an intrinsic correlation between different frequencies, which can vary depending on the ABH and PBH population and their merger rates. Our results demonstrate that the statistical distribution of the SGWB signal can serve as a robust probe of high-redshift stellar properties, which can be explored within the detectable frequency band of the LIGO-Virgo-KAGRA (LVK). The paper is organized as follows: in Sec. \ref{Model}, we discuss the simulation technique developed for different models of the ABHs and PBHs; in Sec. \ref{sum}, we discuss the summary statistics of the SGWB and show the existence of the non-zero off-diagonal terms in the covariance matrix of the SGWB; in Sec. \ref{fis}, we carry out Fisher analysis to obtain the expected constraints on the population parameters from the additional information that can be gained using the off-diagonal terms in the covariance matrix of the SGWB signal; finally, in Sec. \ref{conc}, we discuss the conclusion and the future prospects.

\section{Modelling the contribution from coalescing binaries in the SGWB}\label{Model}

\subsection{Simulation of the SGWB}

We have developed a method to simulate the SGWB signal for different black hole population models. In this section, we briefly describe the method to simulate the SGWB signal.

SGWB due to BBH coalescences is a result of the superposition of individual merger events of all BH properties and formation channels, originating from low to very high redshift. The SGWB can be simulated by simulating individual events of all source masses up to very high redshift, following the given merger rate and mass distribution. In order to simulate the SGWB, we calculate the merger rate as well as mass distribution as a function of redshift. We then divide the redshifts into multiple bins, extending up to a very high redshift. For each bin, we employ Poisson sampling to determine the number of events occurring within a given observation time. For each such event, we sample two component masses from our mass distribution. Finally, we sum the background density due to all events in the given observation time. The simulation is moderately computationally expensive, requiring approximately 90 core hours to generate 1 year of SGWB data with a short-time bin of 200 seconds.

\subsection{Population models for astrophysical black holes}

The ABHs are expected to follow the star formation history. Due to the non-zero time delay between the formation of stars and the merger of the BHs, the merger rate is expected to shift towards lower redshifts compared to the star formation rate (SFR). Here, we model the SFR using the Madau-Dickinson star formation rate \citep{madau2014cosmic}.

\subsubsection{Merger Rate of Astrophysical Black Holes}
The merger rate of ABHs of masses $m_1$ and $m_2$ at redshift z can be written as (\cite{mukherjee2021can,karathanasis2022gwsim})
\begin{equation}
    R_{\rm A}(z,m_1,m_2)= R_{\rm ABH}(z) ~ P_{\rm A}(m_1, z) ~ P_{\rm A}(m_2, z), 
\end{equation}
where $P_{\rm A}(m,z)$ is the mass distribution of the ABHs merging at a redshift of z and $R_{\rm ABH}(z)$ is the source frame merger rate of ABHs per unit comoving volume \citep{karathanasis2022binary}
\begin{equation}
    R_{\rm ABH}(z_m) = R_0 \frac{\int\limits^{\infty}_{z_m} dz \frac{dt}{dz} \times  P_{\rm td}(t(z_m)-t(z)) \times R_{\rm SFR}(z)}{\int\limits^{\infty}_{0} dz \frac{dt}{dz} \times  P_{\rm td}(t(z_m)-t(z)) \times R_{\rm SFR}(z)},
\end{equation}
where $R_0$ is the merger rate at z=0, $R_{\rm{SFR}}$(z) is the Madau-Dickinson star formation rate \citep{madau2014cosmic},

\begin{equation}
    R_{\rm SFR}\propto \frac{(1+z)^{2.7}}{1+(\frac{(1+z)}{2.9})^{5.6}} ,
\end{equation}
and $P_{td}(t)$ is the time-delay distribution \citep{o2010binary,dominik2015double,mandel2016merging,cao2018host,elbert2018counting}, which we model as

\begin{equation}
    P_{\rm td}(t) \propto \bigg\{
    \begin{array}{cl}
    & 0, \quad  t < t_{\rm min}\\
    & t^{-\kappa}, \quad \quad t \geq t_{\rm min} 
    \end{array}
\end{equation}

For a uniform log-space distribution of the initial separation between the binaries, $\kappa=1$ \citep{beniamini2019gravitational,vitale2019measuring,cao2018host}. The current bound on the value of $\kappa$ from GWTC-3 is $\kappa > 0.7$ \citep{karathanasis2022binary}.

\begin{figure}
  \subfigure[]{\label{RA}
    \centering
    \includegraphics[width=\linewidth,trim={0.cm 0  0 0.cm},clip]{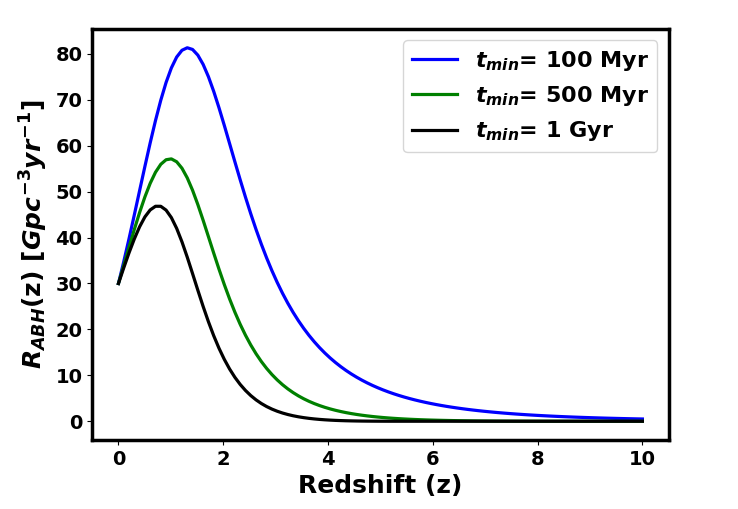}}
  \subfigure[]{\label{PA}
    \centering
    \includegraphics[width=\linewidth,trim={0.cm 0cm  0cm 0.cm},clip]{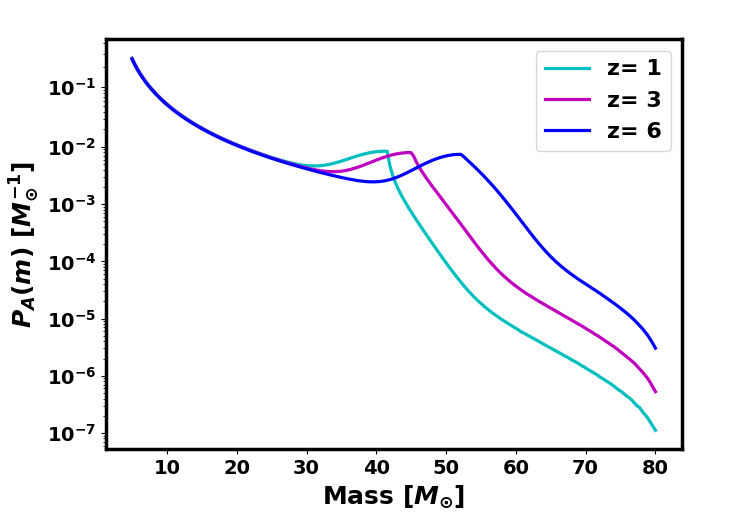}}
  \caption{(a) Merger rate of ABHs for different minimum time-delays, and local merger rate of $30 \, \mathrm{Gpc^{-3} \, yr^{-1}}$. (b) Mass distribution of ABHs merging at different redshifts for minimum delay time, $t_{\mathrm{min}} = 300 \, \mathrm{Myr}$, and $M_{\mathrm{PISN}}(z=0) = 40 \, M_{\odot}$.}
  \label{ABH}
\end{figure}

In Fig. \ref{RA}, we show the merger rate of ABHs with different minimum time-delay ($t_{\rm min}$) for local merger rate of $R_{0}$= 30 $\rm Gpc^{-3}$ $\rm yr^{-1}$. For large $t_{\rm min}$, the peak of the curve shifts towards the lower redshift. This is because
the star formed at some redshift will take a longer time to merge for larger $t_{\rm min}$.

\subsubsection{Mass Distribution of Astrophysical Black Hole}
The upper limit on the mass of the BH is set by the pair-instability supernova  (PISNe) and is known as the PISN limit (around 40 $M_\odot$ to 50 $M_\odot$) \citep{farmer2019mind, heger2002nucleosynthetic,rakavy1967instabilities,fraley1968supernovae,kasen2011pair}. Pair-instability supernova is a type of supernova that occurs in stars with initial masses 140 $M_\odot$ $\lesssim$ M $\lesssim$ 260 $M_\odot$ (final helium core masses of 60 $M_\odot$ $\lesssim$ M $\lesssim$ 140 $M_\odot$) due to pair production (electron-positron pair) in the collision between nuclei and gamma rays produced in the core \citep{,farmer2019mind,kasen2011pair,fryer2001pair}.  PISNe can completely disrupt the star, leaving no stellar remnant behind \citep{gilmer2017pair,farmer2019mind,heger2003massive}.

The PISN mass limit of BHs has been shown, through simulation, to vary with the stellar metallicity \citep{farmer2019mind}. This is due to the mass lost via winds before the star starts pulsating. Stars with high metallicity are expected to lose more mass via wind as compared to low metallicity stars, thus leaving behind a lighter remnant \citep{mokiem2007empirical,van2005metallicity,vink2001mass}. This means that the PISN mass limit is larger for stars with low metallicity. Since the metallicity is known to decrease with redshift, the PISN limit ($M_{\rm PISN}$) is expected to increase with redshift, making mass distribution a redshift-dependent quantity. Therefore, careful modeling of the dependence of $M_{\rm PISN}$ on metallicity can help us understand the metallicity evolution of the universe along with the PISN mass scale itself. We model the metallicity dependence of $M_{\rm PISN}$ as linear in the log of Metallicity (Z) based on the simulations by  \cite{farmer2019mind}.

\begin{equation} \label{Mev}
    M_{\rm PISN}(Z) = M_{\rm PISN}(Z_0) - \alpha ~ \log_{10}(Z/Z_0),
\end{equation}
where $\alpha$ is a parameter that defines the metallicity dependence of $M_{\rm PISN}$ and $Z_0$ is the metallicity at z = 0. Similarly, we model redshift (z) evolution of metallicity (Z) as \citep{mukherjee2022redshift,madau2014cosmic,karathanasis2022binary,karathanasis2022gwsim}

\begin{equation}
    \log_{10}(Z/Z_0) = \gamma ~ z,
\end{equation}
where $\gamma$ is a free parameter.  Combining this with Eq. \eqref{Mev} gives
\begin{equation}
    M_{\rm PISN}(z) = M_{\rm PISN}(0) - \alpha~ \gamma ~ z.
\end{equation}

Due to the time delay between the formation and merger of BHs and redshift-dependent mass distribution, there is going to be a mixing of BHs from different redshifts. The origin of the component BHs can be significantly different which makes the observed mass distribution of BHs very different from the mass distribution of BHs formed at that redshift. We model the mass distribution of the secondary mass of BBHs merging at redshift z by a power law, and the primary mass  by a power law with a bump near $M_{\rm PISN}$, which is
in agreement with the LVK GWTC-3 observation (\cite{abbott2021population}). The bump can arise as a result of the accumulation of the BHs formed from the star undergoing a Pulsation pair-instability supernova (PPISNe) \citep{woosley2017pulsational,farmer2019mind}. The observed mass distribution of the BHs is given by \citep{mukherjee2022redshift, Karathanasis:2022rtr}
\begin{equation}
    P_{A}(z,m) = W(z) \times P_s(z,m), 
\end{equation}
where W(z) is the window function that takes into account the time-delay distribution.
\begin{equation}
    W(z_m) =  \int\limits^{\infty}_{z_m} dz \frac{dt}{dz} \times  P_{td}(t(z_m)-t(z)) \times W_s(z),
\end{equation}
where $W_{s}$(z) is the window function of BH mass at redshift z, and  $P_s(z,m)$ is given by \citep{mukherjee2022redshift, Karathanasis:2022rtr}

\begin{equation}
    P_s(z,m) = (1-\lambda)~ \rm{Pow}(m|a) + \lambda ~  G(m|M_{\rm PISN},\sigma_m),
\end{equation}
where $\mathcal{N}(m| M_{\rm PISN}, \sigma_m)$ is a Gaussian distribution with a mean value $M_{\rm PISN}$ and a standard deviation $\sigma_m$. $\rm Pow(m|a)$ is a power-law distribution with an index a. The parameter $\lambda$ controls the height of the bump. In Fig. \ref{PA}, we show the mass distribution of BHs merging at three different redshifts for $t_{\rm min} = 300$ Myr. The mass distribution shifts towards higher mass at higher redshift.

\begin{table}
\centering
\begin{tabular}{|c|c|}
 \hline
 Parameter & Fiducial value \\ [0.5ex] 
 \hline\hline
 $R_{\rm{ABH}}(0)$ & 30 $\rm{Gpc^{-3} yr^{-1}}$  \\ 
 \hline
 $R_{\rm{PB}H}(0)$ & 20 $\rm{Gpc^{-3} yr^{-1}}$  \\
 \hline
 $t_{\rm{min}}$ & 100 Myr \\
 \hline
 $\alpha$ & 1.5 $M_{\odot}$ \\
 \hline
 $\gamma$ & -1 \\
 \hline
 $\kappa$ & 1  \\ 
 \hline
  $\beta$ & 1.5  \\ 
 \hline
   a & -2.3  \\ 
 \hline
   $\sigma_{p}$ & 0.5  \\ 
 \hline
   $\sigma_{m}$ & 5 $M_{\odot}$  \\ 
 \hline
\end{tabular}
\caption{Table showing the fiducial values of the population parameters.}
\label{tab}
\end{table}

\subsection{Population models of primordial black holes}\label{PBH}
PBHs are hypothesized to have formed in the very early universe soon after the Big Bang \citep{carr2021constraints,carr1975primordial,niemeyer1999dynamics,raidal2017gravitational}. The merger rate of PBHs is expected to increase with redshift \citep{mukherjee2022prospects, mukherjee2021can, ng2022constraining}. We model the merger rate of the PBHs as a power law in redshift \citep{mukherjee2021can, Mukherjee:2021itf}.
\begin{equation}
    R_{\rm{PBH}}(z) = R_{\rm{PBH}}(0) \times (1+z)^{\beta},
\end{equation}
where $R_{\rm{PBH}}(0)$ is the merger rate of the PBHs at z= 0, and $\beta$ is the power-law index . The value of $\beta$ will depend on the amount of clustering exhibited by the PBHs. In the case of Poissonian distribution, the value of $\beta$ $\sim$ 1.3  for most of the scenarios of PBH formation \citep{raidal2017gravitational, sasaki2018primordial, mukherjee2021can}. 

The merger rate along with the mass distribution of the PBHs can shed light on the properties and characteristics of dark matter \citep{bird2022snowmass2021,carr2016primordial,belotsky2014signatures}. It is believed that the PBHs may constitute a notable portion of the dark matter, and their population and mass distribution can help us constrain the properties of dark matter. we characterize the mass distribution of the PBHs as log-normal distribution \citep{dolgov1993baryon, carr2017primordial}.

\begin{equation}
    P_{PBH} = \frac{1}{\sqrt{2 \pi} \sigma_p m} \times \exp[-\frac{(\log(m/M_c))^2}{2 \sigma_{p}^{2}}],
\end{equation}
where $M_c$ is the characteristic mass scale and $\sigma_{p}$ is the standard deviation of the $\log(m/M_c)$. In Table. \ref{tab}, we list the fiducial values of the population parameters considered in the paper. 

\begin{figure}
    \centering
    \includegraphics[width=8.3cm]{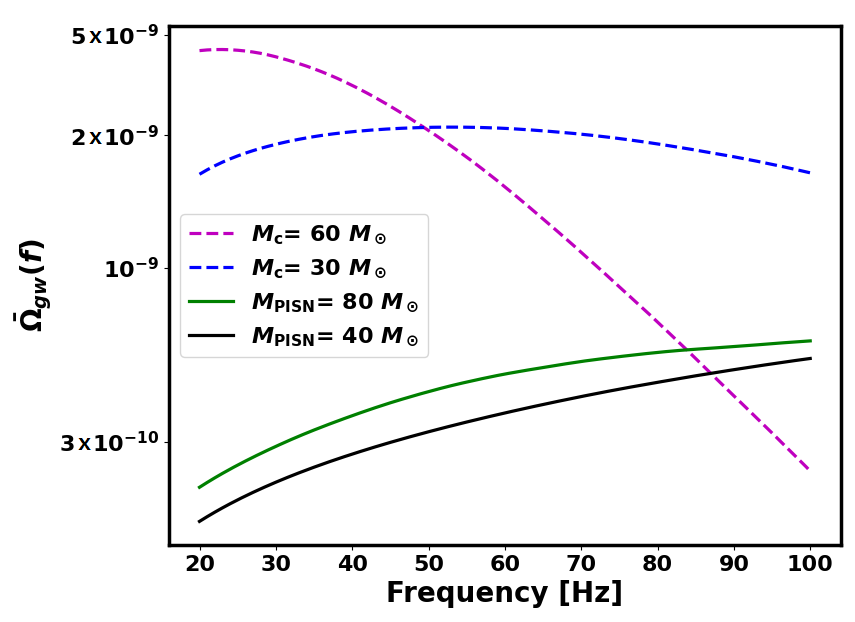}
  
    \caption{SGWB density due to ABHs for different values of the PISN mass scale $M_{\rm PISN}$ (solid lines) and due to PBHs for different values of the characteristic mass scale $M_c$ (dashed lines). The local merger rate of ABHs and PBHs is taken as 30 $\rm Gpc^{-3} \rm yr^{-1}$ and 20 $\rm Gpc^{-3} \rm yr^{-1}$ respectively.}
    \label{BG1}
\end{figure}

\section{Summary statistics of the SGWB from simulations}\label{sum}
We use simulation-based techniques to estimate the power spectrum of the SGWB, as well as the non-stationary behaviour of the SGWB. We show below the results from our simulations of the (i) SGWB power spectrum and (ii) the signal covariance matrix of the SGWB, and the effect of non-stationary signals on the correlation between different frequency modes.

\begin{figure}
    \centering
    \includegraphics[width=8.5cm]{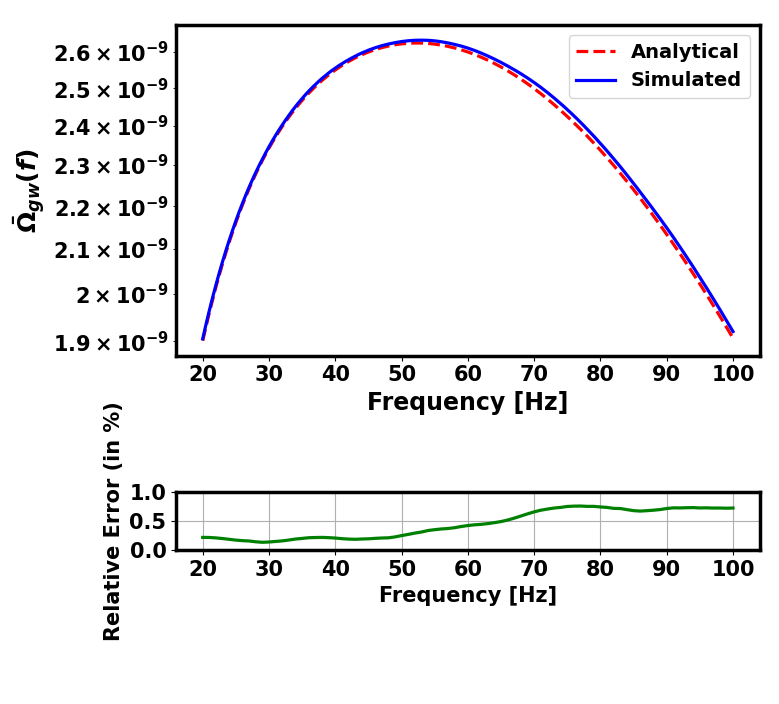}
  
    \caption{Analytically obtained $\overline{\Omega}_{gw}(f)$ compared to simulated  $\overline{\Omega}_{gw}(f)$ (top) due to PBHs and the relative difference between them (bottom), for local merger rate of  20 $\rm Gpc^{-3} \rm yr^{-1}$ and $M_c$= 30 $M_\odot$.}
    \label{Sim}
\end{figure}
\subsection{SGWB power spectrum}

The SGWB is defined as the energy density per logarithmic frequency interval divided by
the critical energy density ($\rho_c$ $c^2$) required to close the universe. The SGWB density can be written as 
\citep{phinney2001practical,christensen2018stochastic}

{
\begin{equation}
    \begin{aligned}
     \overline{\Omega}_{\rm gw}(f) = \frac{1}{\rho_c c^2} &\int\limits^{m_{max}}_{m_{\rm min}} dm_1 \int\limits^{m_{max}}_{m_{\rm min}} dm_2 \int\limits_{z_{\rm min}}^{\infty} f_r ~dz~ \frac{dV_c}{dz} \\
    &\times \bigg[\frac{R_{\rm{GW}}(z,m_1,m_2)}{1+z}\bigg]  \bigg[\frac{1+z}{4 \pi d_{L}^{2} c} \frac{dE_{\rm{gw}}}{df_r} \bigg],
    \end{aligned}
    \label{SGWB}
\end{equation}
}
where $R_{\rm{GW}}(z,m_1,m_2)$ is the source frame merger rate of BHs per unit comoving volume between masses $m_1$ and $m_2$ at redshift z, $d_L$ is the luminosity distance of the source, and $f_r$ = $f$(1+z) is the source frame frequency. We take $M_{\rm{min}}$= 5 $M_{\odot}$, and the value of $z_{\rm{min}}$ is selected as the maximum redshift at which the A+ detector is sensitive to the most probable mass of the considered mass distribution with the most probable value of the orientation parameter '$\Theta$' \citep{finn1993observing}. $\frac{dE_{\rm{gw}}}{df_r}$ is the energy emitted by the source per unit source frame frequency ($f_r$), 
\begin{equation}
    \frac{dE_{gw}}{df_r} = \frac{(G \pi)^{2/3}  M_c^{5/3} \times \Pi(f_r)}{3},
\end{equation}
where $M_c$ is the chirp mass of the source, and $\Pi(f_r)$ is given by \citep{ajith2008template}
    \begin{equation}
    \Pi(f_r) =\left\{
    \begin{array}{c l}
    & f_r^{-1/3}, \quad \quad f_r < f_{merg}\\
    &\frac{f_r^{2/3}}{f_{merg}}, \quad \quad    f_{merg} \leq f_r < f_{ring}\\
    &\frac{1}{f_{merg}~ f_{ring}^{4/3}} \left(\frac{f_r}{1+(\frac{f_r-f_{ring}}{f_{\omega}/2})^2}\right)^2, f_{ring} \leq f_{r} < f_{cut},
    \end{array}
    \right.
\end{equation}
where, $f_{merg}$, $f_{ring}$, and $f_{cut}$  are cut-off frequencies for inspiral, merger, and ringdown stages respectively.

In Fig \ref{BG1}, we show how the SGWB power spectrum is impacted by the different mass scales of the BH population of the astrophysical and primordial origin in solid lines and dashed lines respectively. It can be observed that the population with higher masses exhibits a larger power density at low frequencies. This is because the presence of a Gaussian peak in the mass distribution of the BHs at M$_{\rm PISN}$ predominantly leads to more contribution at the lower frequencies of the SGWB.

The presence of PBHs introduces a distinctive signature in SGWB. As the merger rate of the PBHs is expected to increase with redshift, the dominant contribution to SGWB due to PBHs will come from higher redshifts. The gravitational waves from sources at higher redshifts are going to be redshifted towards lower frequencies. This means the presence of PBHs will modify the shape of the SGWB spectrum such that we have enhanced power at low frequencies.

In Fig. \ref{Sim}, we compare the analytically obtained $\overline{\Omega}_{gw}(f)$ with the $\overline{\Omega}_{gw}(f)$ obtained using the simulation developed in this work. There is less than $1\%$
discrepancy between the two results.

\subsection{Non-stationary SGWB: off-diagonal terms in the covariance matrix}
The SGWB will not be uniform over time; rather, it is expected to exhibit temporal fluctuation as discussed in Sec. \ref{intro}. This fluctuation arises due to the limited number of events and the diverse properties of the gravitational wave sources. This makes SGWB a non-stationary quantity. However, the SGWB signal will exhibit time translation symmetry over a large timescale, as the astrophysical source population will remain constant over the observation time scale. So, it is important to understand at what timescales the signal homogenizes and becomes statistically stationary (time-translation symmetric). The presence of a non-stationary signal will lead to the off-diagonal terms in the covariance matrix in the frequency domain. We will discuss this below in detail for different population models.

The probability distribution of the number $N$ of events in an interval $\Delta T$ is given by 
\begin{equation}
    P(N) = \frac{(R\Delta T)^{N}}{N!} \times e^{-(R \Delta T)}, 
\end{equation}
where $R$ is the mean merger rate. The standard deviation in the number of events is given by (R $\Delta
T)^{1/2}$ \citep{dvorkin2018exploring,bulik2011ic10,kalogera2007formation}. Therefore the relative fluctuation will scale as (R $\Delta
T)^{-1/2}$. This means there is going to be significant fluctuation in the SGWB if the merger rate is below a certain value. The distribution of SGWB and associated statistical quantities can be shown to vary with BH population parameters leading to novel non-Gaussian signatures. The fluctuation in SGWB density $\Delta$$\Omega_{gw}$(f) can be defined as 
\begin{equation}
    \Delta\Omega_{\rm gw}(f) = \sqrt{\left<\left({\frac{\Omega_{\rm gw}(f,t) - \overline{\Omega}_{\rm gw}(f)}{\overline{\Omega}_{\rm gw}(f)}}\right)^2\right>},
\end{equation}

The fluctuation $\Delta\Omega_{\rm gw}(f)$ is going to be small for large merger rates. We can define three-time scales \citep{mukherjee2020time, Mukherjee:2020jxa}: $\tau$, which is the duration of the signal for an event, $\Delta t_{\rm{event}}$, which is the mean duration between two consecutive events, and $\Delta T$, which is the observation time bin. The $\tau$ depends on the chirp mass of the source ($\tau \propto M_{c}^{-5/3}$). Typically, for BBHs $\tau$ can vary from a fraction of a second to a few seconds. The time scale $\Delta t_{\rm{event}}$ depends on the merger rate. For a large merger rate, the interval between two events is going to be small. Hence, we are going to have overlapping events for $\Delta t_{\rm{event}}  < \tau$. The fluctuation of $\Omega_{gw}(f)$ under this condition is going to be very low. Therefore, we are going to encounter a fluctuating SGWB only under condition $ \tau < \Delta t_{\rm{event}}$.
For a local merger rate of 30 $\rm Gpc^{-3}$  $\rm yr^{-1}$ and $t_{\rm min}$ = 100 Myr, $\Delta t_{\rm event}$ $\sim$ 1000 seconds (for the case with only ABHs).

\subsubsection{Model dependence of SGWB distribution}
The fluctuation and distribution of the SGWB depend on the population parameters of BHs. The time dependence of the background signal can serve as additional information on the top of the power spectrum, to infer various population parameters of BHs, like merger rate and mass distribution, as well as to distinguish different formation channels such as primordial and astrophysical \citep{mukherjee2020time}. 

\begin{figure}
    \centering
     \subfigure[]{\label{DistMp}
    \includegraphics[width=\linewidth,trim={0.cm 0  0 0.cm},clip]{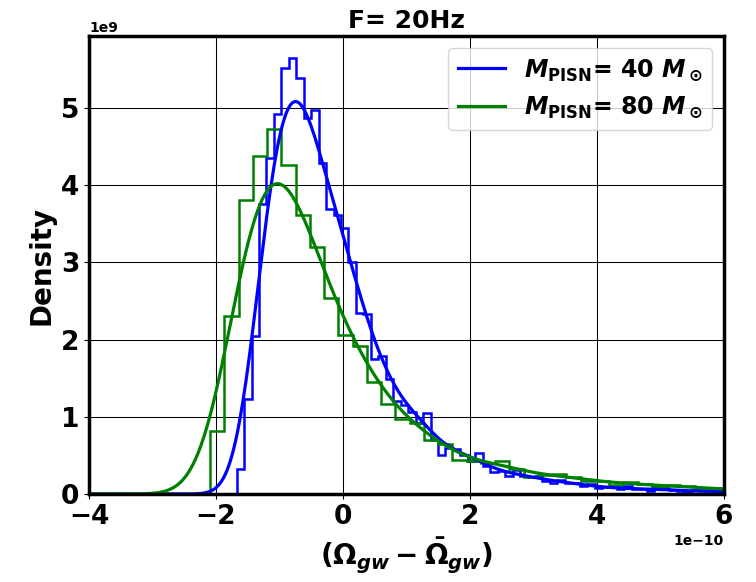}}
    
  \subfigure[]{\label{DistMc}
    \centering     \includegraphics[width=\linewidth,trim={0.cm 0  0 0.cm},clip]{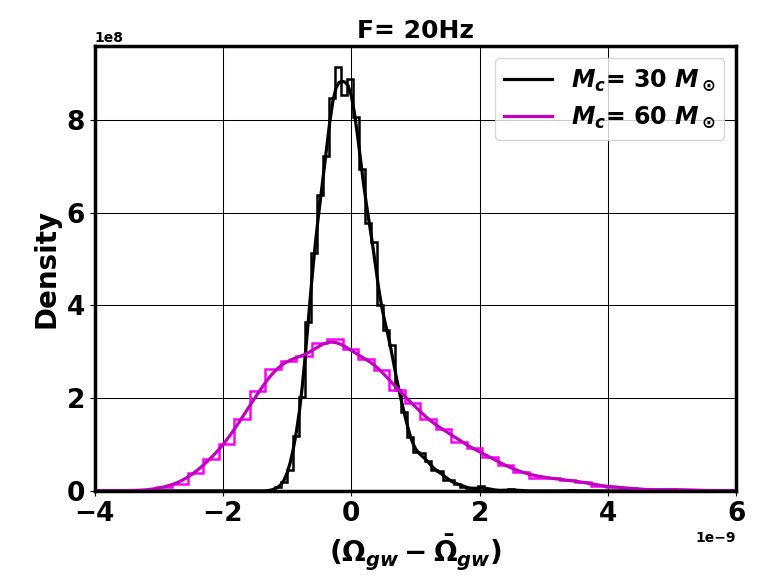}}
    \caption{Distribution of fluctuation in ${\Omega}_{gw}(f)$ due to (a) ABHs for different $M_{\rm{PISN}}$ (b) PBHs for different $M_{\rm{c}}$, for a short-time bin of $10^{4}$ seconds.}
\label{DistM}
\end{figure}

\begin{figure}
    \centering
    \subfigure[]{\label{DistMc20}
    \includegraphics[width=\linewidth,trim={0.cm 0  0 0.cm},clip]{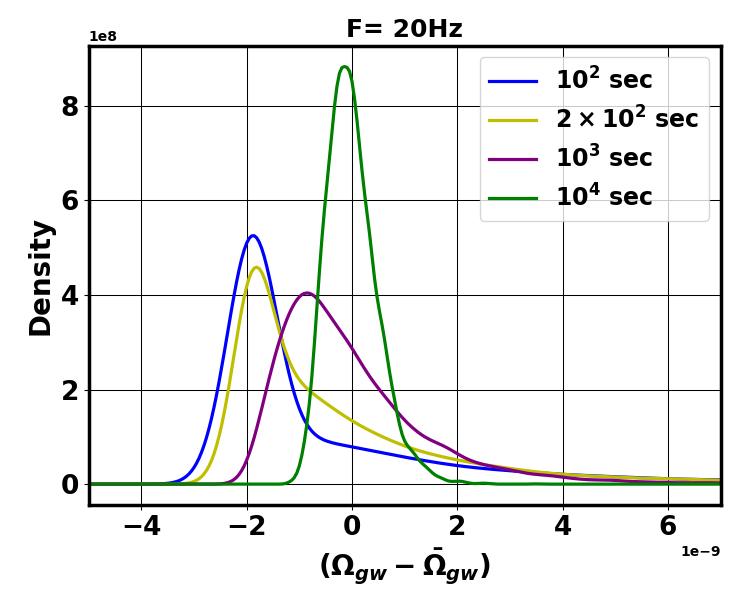}}
   \subfigure[]{\label{DistMc60}
    \centering \includegraphics[width=\linewidth,trim={0.cm 0  0 0.cm},clip]{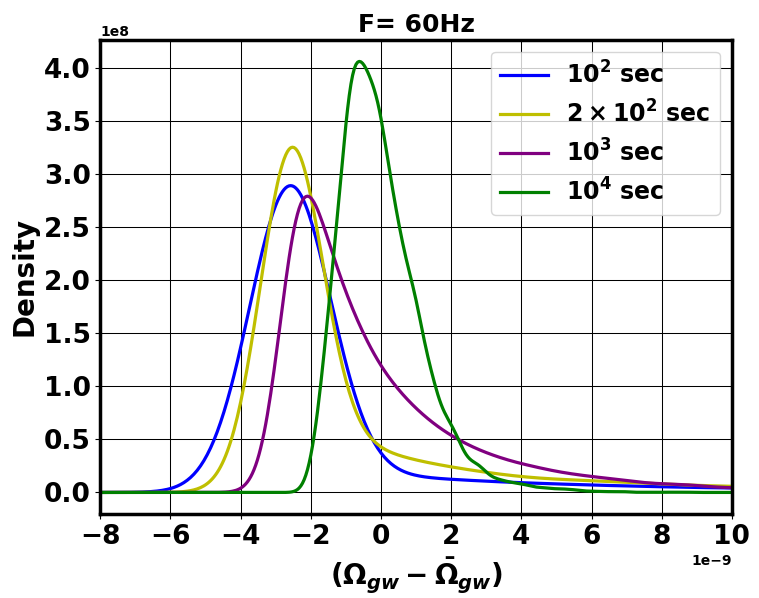}}
    \caption{Distribution of fluctuation in ${\Omega}_{gw}(f)$ due to PBHs for $M_c$= 30 $M_\odot$ and $\beta$= 1.5 at (a) F= 20 Hz, (b) F= 60 Hz, for different short-time bins.}
\label{DistMcdt}
\end{figure}

We show in Fig. \ref{DistMp} the distribution of fluctuation in $ \Omega_{\rm gw}(20~\rm{Hz})$ for ABHs for exposure of $10^4$ seconds, for $M_{\rm PISN}$= 40 $M_\odot$ and $M_{\rm PISN}$= 80 $M_\odot$. It can be seen that higher mass makes the distribution of SGWB relatively more dispersed and skewed at $f=20$ Hz. This is because a higher mass source generates larger strain at the low frequencies. In general, the skewness of the distribution of the background signal at different frequencies depends on the underlying population of the masses contributing to the background. A large skewness also implies there are a relatively larger number of realizations with no events.
In Fig. \ref{DistMc}, we also show the distribution of fluctuation in $ \Omega_{\rm gw}(20~\rm{Hz})$ for PBHs for $M_{\rm {c}}$= 30 $M_\odot$ and $M_{\rm {c}}$= 60 $M_\odot$.

In Fig. \ref{DistMcdt}, we show the distribution of the fluctuation for different observation time bins over which the signal is averaged. The distribution becomes more and more skewed as we decrease the bin size. For bins smaller than $10^{4}$ seconds, we have a significant number of realizations where we do not have any events. The distribution of $\Omega_{gw}(f)$, therefore, stacks at zero, and the distribution of fluctuation in $\Omega_{gw}(f)$ shifts towards negative value. The key signature here is that a higher merger rate and a larger time bin will make the distribution Gaussian. Also, the change in the mass distribution will change the skewness of the distribution, as shown in Fig. \ref{DistM}. Therefore, the fluctuation in SGWB is going to be very critical in understanding the merger rate, mass distribution, and other population parameters including the formation channels of binaries.

\subsubsection{Model dependence of SGWB signal covariance matrix}
Apart from the fluctuation and skewness of the SGWB distribution, the SGWB is expected to exhibit correlations between signals at different frequencies. These correlations arise due to the non-stationary nature of the GW signal contributing to the background. In the case of a low merger rate, there will always be a non-zero degree of correlation between signals at different frequencies within the time scale over which the signal homogenizes. This correlation is expected to depend on population parameters, particularly the mass distribution and the merger rate. The presence of the correlation between signals at different frequencies will lead to a non-zero off-diagonal covariance matrix. Here, we provide a summary of how the signal covariance matrix is influenced by the properties of the BH population namely,  (i) mass distribution and (ii) merger rate. We have restricted our analysis up to $f=100$ Hz, as the signal beyond this is not well measured due to small values of the overlap reduction function \citep{PhysRevD.48.2389, PhysRevD.46.5250}. 

\begin{figure}
    \centering
   \subfigure[]{\label{CovMp40}
    \includegraphics[width=\linewidth]{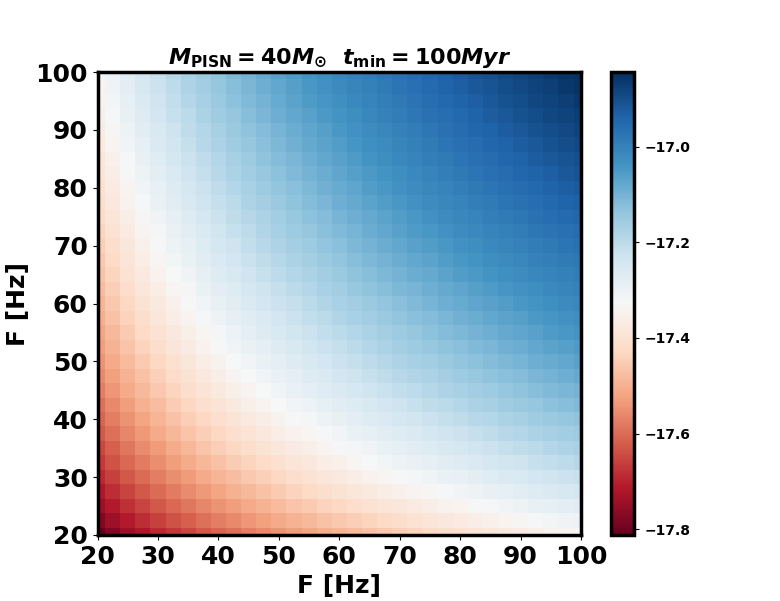}}
    \subfigure[]{\label{CovMp80}
    \includegraphics[width=\linewidth]{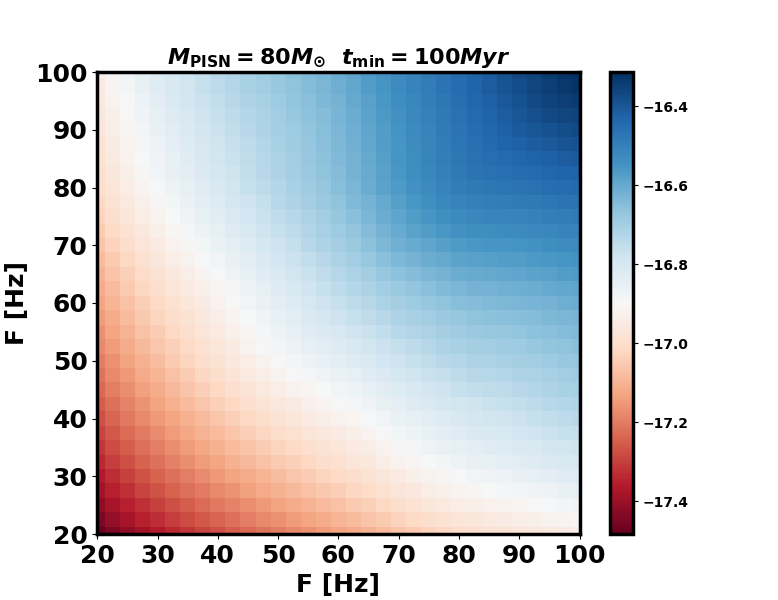}}
    \caption{Covariance matrix of $\Omega_{gw}(f)$ (see eq. \ref{Covs}), for short time bin $\Delta T = 200$ seconds, due to ABHs showing covariance between different frequency modes for (a) $M_{\rm PISN}$= 40 $M_\odot$, (b) $M_{\rm PISN}$= 80 $M_\odot$, and $t_{\rm min}$= 100 Myr, in logarithmic scale.}
\label{CovMp}   
\end{figure}

\begin{figure}
    \centering
    \subfigure[]{\label{CovMc30}
    \includegraphics[width=\linewidth]{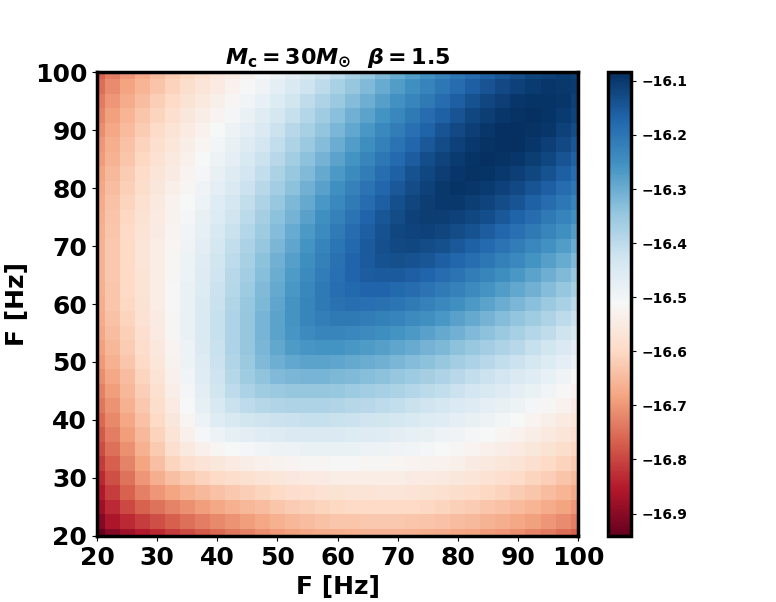}}
    
    \subfigure[]{\label{CovMc60}
    \centering
    \includegraphics[width=\linewidth]{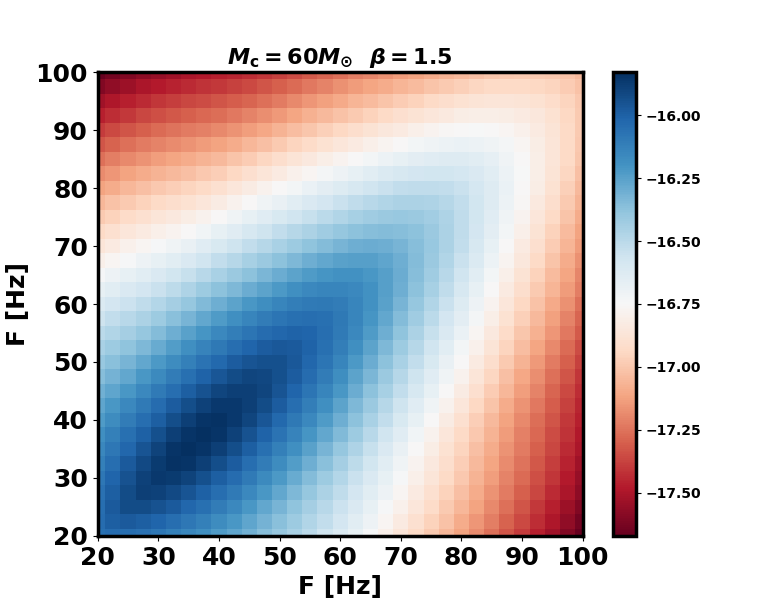}}
    \caption{Covariance matrix of $\Omega_{gw}(f)$ (see eq. \ref{Covs}), for short time bin $\Delta T = 200$ seconds, due to PBHs showing covariance between different frequency modes for (a) $M_c$= 30 $M_\odot$, (b) $M_c$= 60 $M_\odot$, and $\beta$= 1.5, in logarithmic scale.}
\label{CovMc}    
\end{figure}

\textit{\textbf{Impact of mass distribution on covariance matrix:}}
If a mass distribution has lower mass BHs, we expect to see the correlation between the signals at different frequencies up to larger frequency separation as low-mass BHs emit across a broader frequency range. Likewise, the mass distribution with a small variance is also expected to show more correlation as compared to the case with a large variance in mass distribution because the sources with different masses have distinct power spectra. Therefore, the mass distribution significantly influences the overall structure of the covariance matrix. In Fig. \ref{CovMp} and Fig. \ref{CovMc}, we show the covariance matrix of $\Omega_{gw}(f)$ for a short-time bin width of 200 seconds (over which
the signal is averaged) for ABHs and PBHs respectively. For a short-time bin width of 200 seconds, only a fraction of realizations have an event. In such a case, the signals at different frequencies are highly correlated. Since the maximum frequency that sources can emit is inversely proportional to its mass, the signal covariance descends more rapidly as we move away from the diagonal term in the covariance matrix for the case with a higher mass population. Similarly, the peak in the covariance matrix shifts towards the lower frequencies for distribution with higher mass BHs because the higher mass BHs emit mostly at lower frequencies. 

\textit{\textbf{Impact of merger rate on covariance matrix:} }

The merger rate has a direct impact on the magnitude of the covariance matrix. A higher merger rate corresponds to a shorter time interval, $\Delta t _{\rm{event}}$, between two events. The decrease in $\Delta t_{\rm{event}}$ leads to a smaller fluctuation in SGWB, resulting in a more Gaussian distribution. However, the merger rate does not alter the shape of the covariance matrix; instead, it scales the overall magnitude of the covariance matrix. As a result, it becomes possible to explore both the mass distribution and merger rate of the sources contributing to the SGWB signal using the power spectrum and the covariance matrix.

\section{Fisher Forecast}\label{fis}
In this section, we perform a Fisher analysis to show the advantage of the additional information from the covariance matrix in constraining the population parameters. The Fisher matrix analysis is a powerful statistical tool used to estimate how well a set of model parameters can be constrained based on a given set of data \citep{fisher1935logic,tegmark1997karhunen}. The Fisher information matrix is given by

 \begin{equation}
     F_{ij}= 
 \Big<\frac{\partial^{2} \mathcal{L} }{\partial \theta_{i} \partial \theta_{j}}\Big>,
\label{Fisher}
\end{equation}

 where $\mathcal{L}= -\ln$ $L$, where $L$ is the likelihood function, and $\theta_{i}$ and $\theta_{j}$ are model parameters. According to the Cram\'er-Rao inequality \citep{rao1945information,cramer1946contribution}, the minimum uncertainty in the measurement of a model parameter is given by $\Delta \theta_{i} \geq 1/\sqrt{F_{ii}}$ 
(\cite{tegmark1997karhunen}). For Gaussian likelihood, 

\begin{equation}
    \begin{aligned}
    \ln[L(\hat{\Omega}_{\rm gw}|\vec{\Theta})] \propto \sum\limits_{k}\Big[\ln(\det \mathbf{C}) & +  \big[\hat{\mathbf{\Omega}}_{\rm gw}^{k} - \mathbf{\overline{\Omega}}_{\rm gw}^{m}\big] \\
    & \mathbf{C}^{-1} \big[\mathbf{\hat{\Omega}}_{\rm gw}^{k} - \mathbf{\overline{\Omega}}_{\rm gw}^{m}\big]^{T} \Big],
    \end{aligned}
\end{equation}
where $\vec{\Theta} \in\{\theta_1, \theta_2, \hdots, \theta_n\}$ represents population parameters, $\hat{\Omega}_{\rm gw}^{k}(f)$ is the measured SGWB density signal from $k^{th}$ short-time bin, $\overline{\Omega}_{\rm gw}^{m}(f)$ is the mean value of SGWB signal for parameters $\vec{\Theta}$, and $\mathbf{C}$ is the covariance matrix for signal averaged over given short-time bin. The covariance matrix $\mathbf{C}$ is given by
\begin{equation}
    \begin{aligned}
    \rm{C}(f,f') =&  \Big<\big(\hat{\Omega}_{\rm gw}(f) - \overline{\Omega}_{\rm gw}^{m}(f)\big)
     \big(\hat{\Omega}_{\rm gw}(f') -\overline{\Omega}_{\rm gw}^{m}(f')\big) \Big>\\
     =&~ \rm{C}_{N} (f,f') + \rm{C}_{S} (f,f'),
    \end{aligned}
    \label{Cov1}
\end{equation}
where $\rm{C}_{N} (f,f')$ is the noise
covariance matrix \citep{thrane2009probing,christensen2018stochastic}, and $\rm{C}_{S} (f,f')$ is the covariance due to intrinsic fluctuation in ${\Omega}_{\rm gw}(f)$ that can be written as

\begin{equation}
     \rm{C}_{S}(f,f')= \Big<(\Omega_{gw}^{m}(f)-\overline{\Omega}_{gw}^{m}(f)) (\Omega_{gw}^{m}(f')-\overline{\Omega}_{gw}^{m}(f'))\Big>,
     \label{Covs}
\end{equation}
where $\Omega_{gw}^{m}(f)$ is the signal averaged over given short-time bin $\Delta T$. The intrinsic covariance of $\Omega_{gw}(f)$ can be useful as it provides additional information that can help us better constrain the population parameters. All the analyses of the SGWB power spectrum have disregarded the correlation between frequency bins. We demonstrate the advantage of the full covariance matrix and the impact of its off-diagonal terms on estimating source properties.

The noise covariance matrix, in this analysis, is assumed to be diagonal. The diagonal assumption is a good approximation for stationary noise (\cite{abbott2020guide}). However, in the non-stationary noise limit, we can expect to see the spectral correlation in the noise power spectrum \citep{abbott2020guide,nuttall2018characterizing,mozzon2022does}. The time scale over which the noise is stationary is much larger than the time scale of the BBH signals. As a result, the off-diagonal terms in the signal covariance matrix remain uncontaminated from the noise. Also in the presence of non-stationary noise (such as glitches), it will be uncorrelated between the different detectors (except for the Schumann resonance \citep{schumannUberDampfungElektromagnetischen1952,nguyenEnvironmentalNoiseAdvanced2021,thraneCorrelatedMagneticNoise2013,
thraneCorrelatedNoiseNetworks2014}). The spectrum of the Schumann resonance is different from the astrophysical SGWB signal. In this analysis, we assume the noise to be Gaussian and stationary. In future work with the LVK data, we will explore the contribution of the non-stationary noise.

The  Fisher information matrix in Eq. \eqref{Fisher} can be expanded as (\cite{fisher1935logic, tegmark1997karhunen})
\begin{equation}
    \begin{aligned}
    F_{ij}= &
    \frac{T_{\rm obs}}{2~\Delta T} \times \rm{Tr}\Big[(\mathbf{C})^{-1} \mathbf{C}_{,i}~ \mathbf{C}^{-1} \mathbf{C}_{,j}+ \mathbf{C}^{-1}~\mathbf{A}_{ij} \Big],
    \end{aligned}
 \end{equation}
where $T_{\rm obs}$ is the total observation time, $\Delta T$ is the short-time bin, $\mathbf{C}_{, i}$ denotes the derivative of the covariance matrix with respect to the parameters $\theta_i$, and $\mathbf{A}_{ij}$= [($\overline{\mathbf{\Omega}}_{\rm gw}^{m})_{,i}$  ($\overline{\mathbf{\Omega}}_{\rm gw}^{m})_{,j}^{T}$ + ($\overline{\mathbf{\Omega}}_{\rm gw}^{m})_{,j}$  ($\overline{\mathbf{\Omega}}_{\rm gw}^{m})_{,i}^{T}$].

\begin{figure}
    \centering
     \subfigure[]{\label{gtcMp40}
\includegraphics[width=\linewidth,trim={0.cm 0cm  0cm 0.cm},clip]{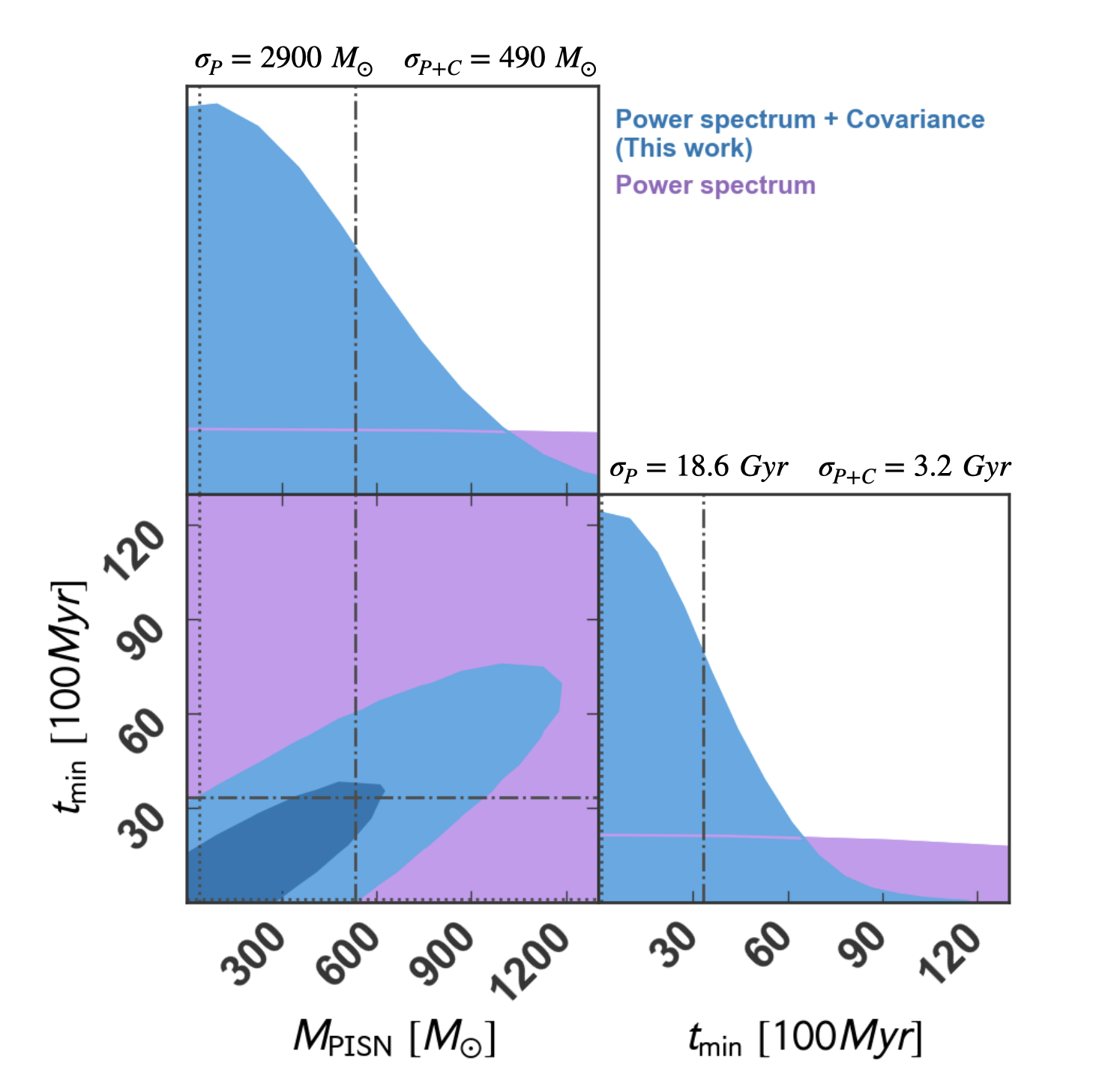}}
     \subfigure[]{\label{gtc_Mp80}
    \centering
\includegraphics[width=\linewidth,trim={0.cm 0cm  0cm 0.cm},clip]{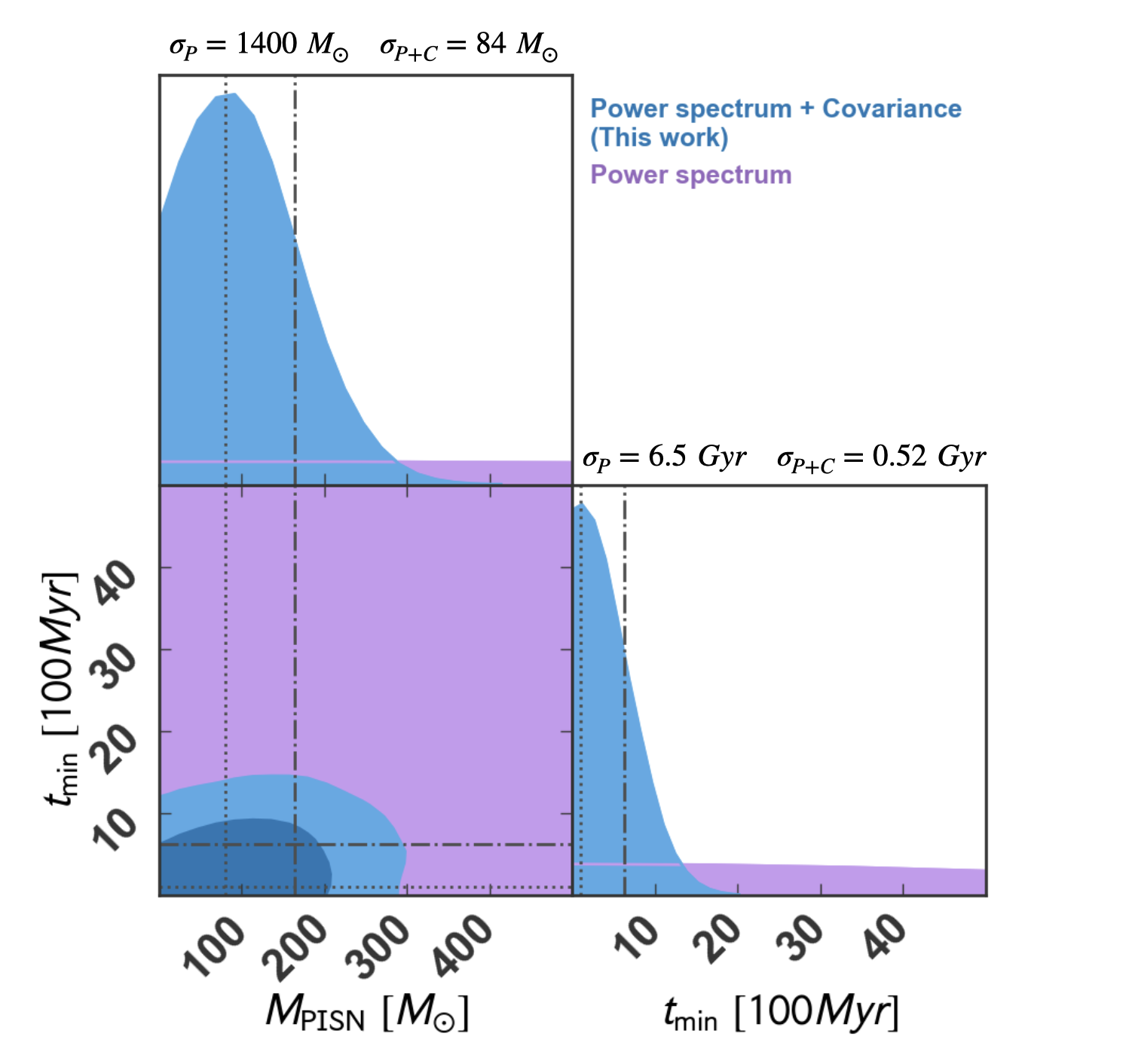}}
    \caption{Corner plot showing the feasibility of measuring the $M_{\rm{PISN}}$ and $t_{min}$ for (a) $M_{\rm{PISN}}$= 40 $M_\odot$, (b) $M_{\rm{PISN}}$= 80 $M_\odot$, and $t_{min}$= 100 Myr for the ABH population. The purple color represents the case where we have assumed a stationary background and the blue color represents the case where we have included the intrinsic covariance of the SGWB.}\label{gtcMp}
\end{figure}

\begin{figure}
    \centering
   \subfigure[]{\label{gtcMc40}
    \centering
    \includegraphics[width=\linewidth,trim={0.cm 0  0 0.cm},clip]{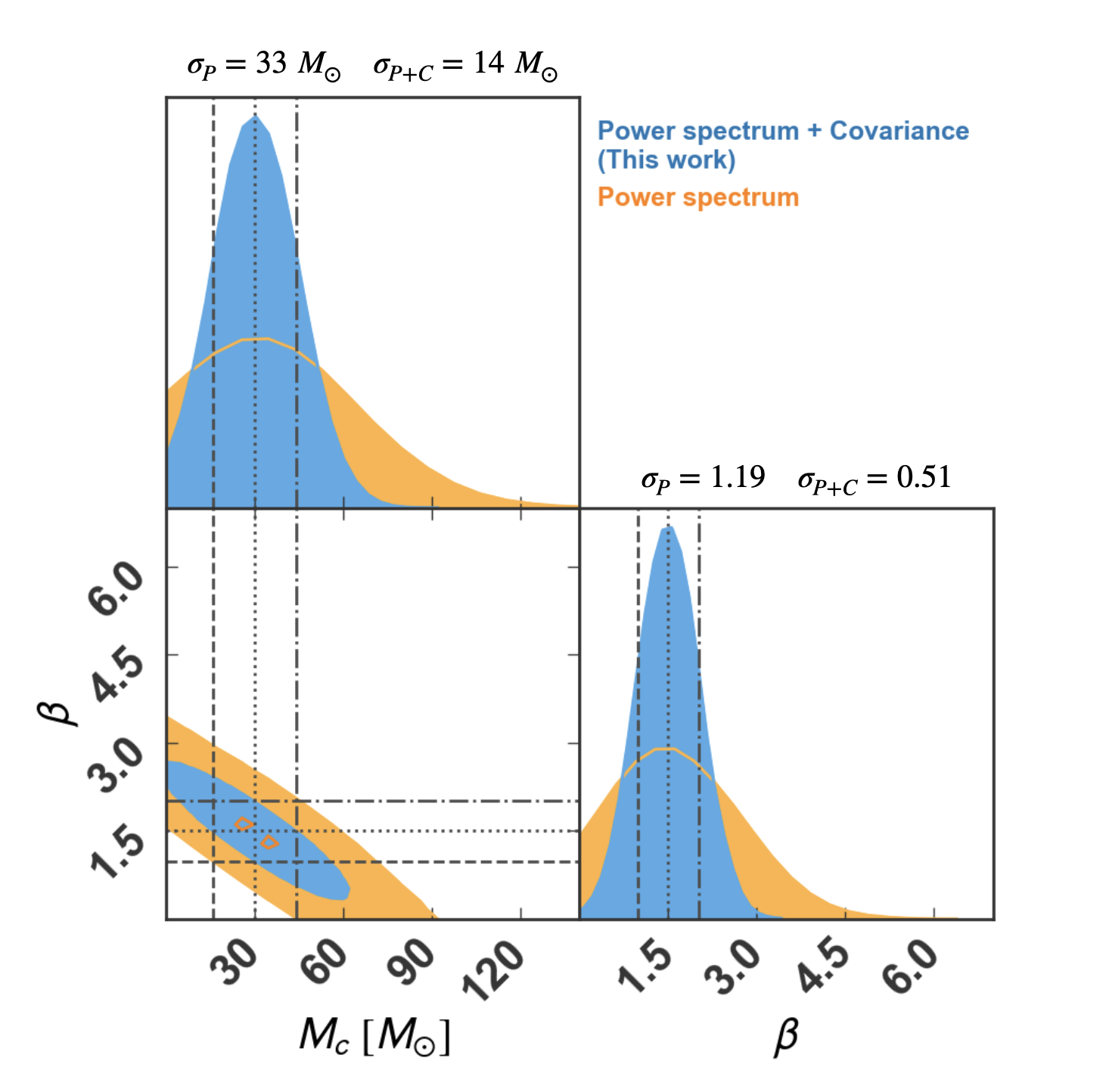}}
     \subfigure[]{\label{gtcMc80}
    \centering
    \includegraphics[width=\linewidth,trim={0.cm 0  0 0.cm},clip]{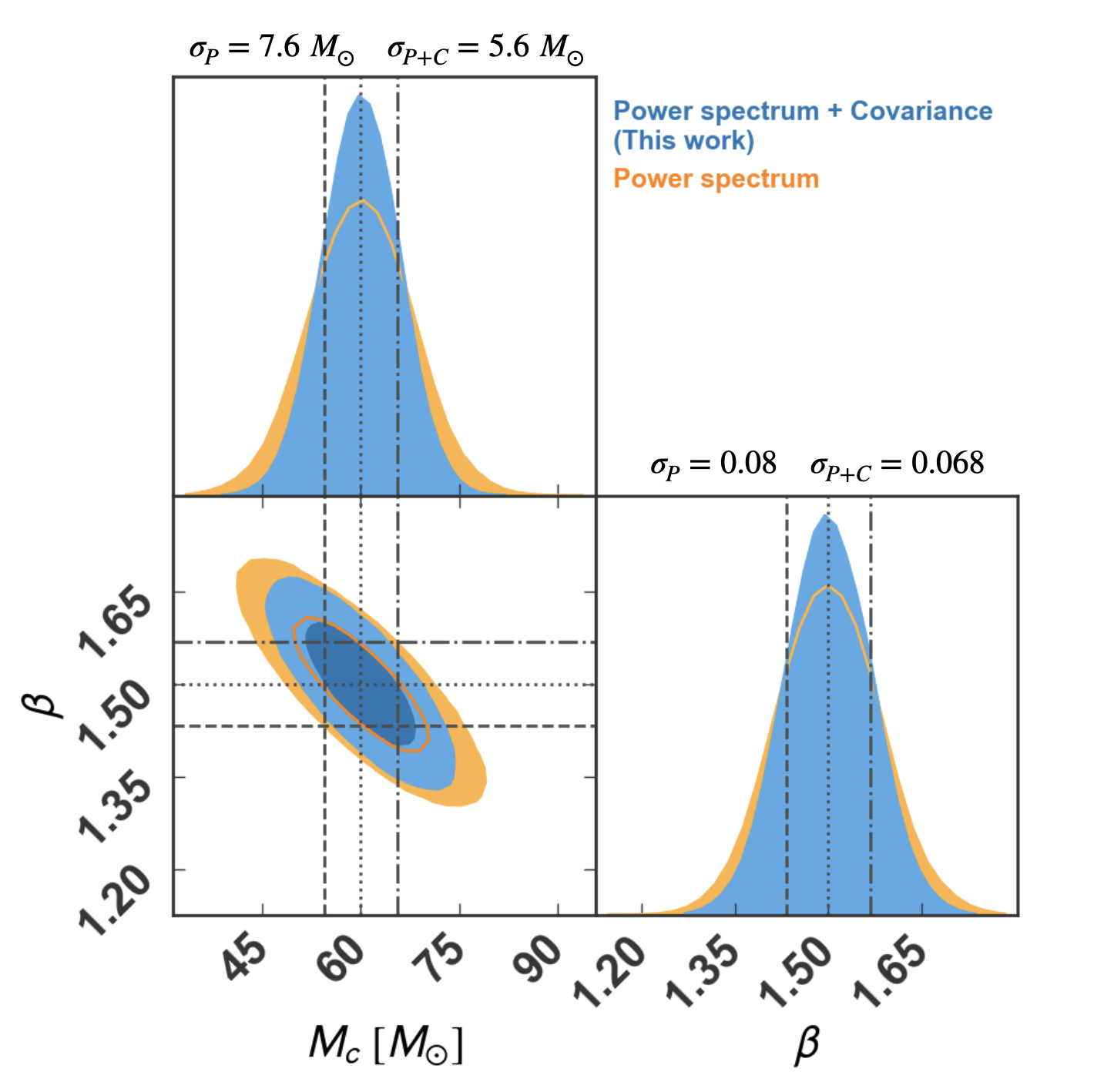}}
    \caption{Corner plot showing the feasibility of measuring $M_c$ and $\beta$ for (a) $M_c$= 30 $M_\odot$, (b) $M_c$= 60 $M_\odot$, and $\beta$= 1.5 for the PBH population. The orange color represents the case where we have assumed a stationary background and the blue color represents the case where we have included the intrinsic covariance of the SGWB.}
\label{gtcMc}
\end{figure}

 The covariance matrix, $\mathbf{C}$ in Eq. \eqref{Cov1}, has contributions from two sources: (i) instrument noise and (ii) intrinsic fluctuations in $\Omega_{gw}(f)$. We perform the Fisher analysis on the $\Omega_{gw}(f)$ signal to obtain the expected constraints on the parameters. We consider two cases: (A) where we assume the signal to be stationary with no intrinsic fluctuation (power spectrum-only case), and (B) where we also include the covariance due to intrinsic fluctuation (power spectrum + covariance). We apply this analysis to the A+ sensitivity of the LIGO (Hanford \& Livingston) and Virgo detectors  \citep{Aasi:2013wya, TheLIGOScientific:2014jea, TheVirgo:2014hva, barsotti2018a+, aplus} for a total observation period of two years. In case (B), we have considered a short-time bin size of 200 seconds. In Fig.  \ref{gtcMp} and Fig. \ref{gtcMc}, we show the Gaussian Fisher posterior distribution for the above two cases. We find that for ABHs, the inclusion of the covariance matrix can improve constraints on both parameters, $M_{\rm{PISN}}$ and $t_{\rm{min}}$, by up to about 90\%. For PBHs, the improvement is more significant for $M_c$= 30 $M_{\odot}$, with an enhancement in the measurement by approximately 55\%, compared to $M_c$ = 60 $M_{\odot}$, where it is only around 17\%.

\begin{figure}
    \centering
\includegraphics[width=0.5\textwidth]{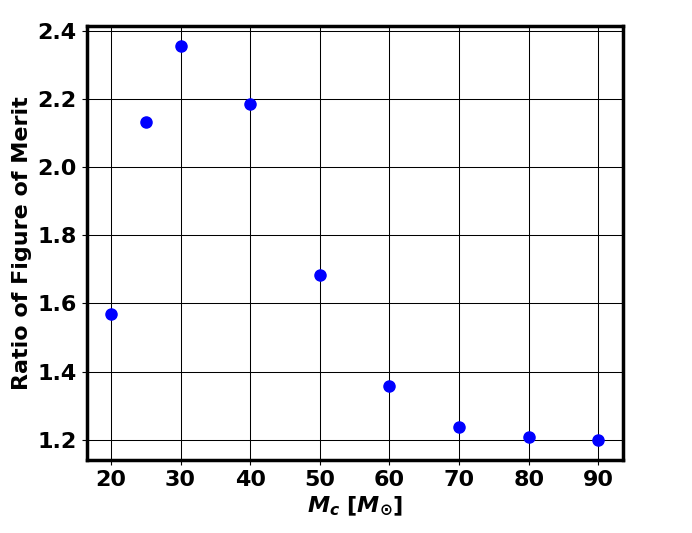}
    \caption{The ratio of Figure of Merit (FoM) for the
    power spectrum + covariance case to that for the power spectrum only case of PBHs, as a function of $M_c$ for a short-time bin of $\Delta T= 200$ seconds.}
    \label{FoM-M}
\end{figure}

\begin{figure}
    \centering
    \includegraphics[width=0.5\textwidth]{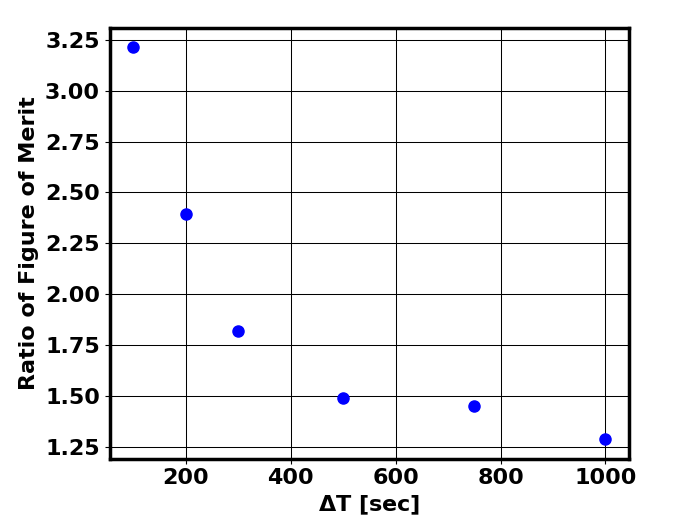}
    \caption{The ratio of Figure of Merit (FoM) for the
    power spectrum + covariance case to that for the power spectrum-only case, as a function of the short-time bin ($\Delta$T) for PBHs with $M_c$= 30 $M_\odot$.}
    \label{FoM-T}
\end{figure}

\begin{figure}
    \centering
\includegraphics[width=0.5\textwidth]{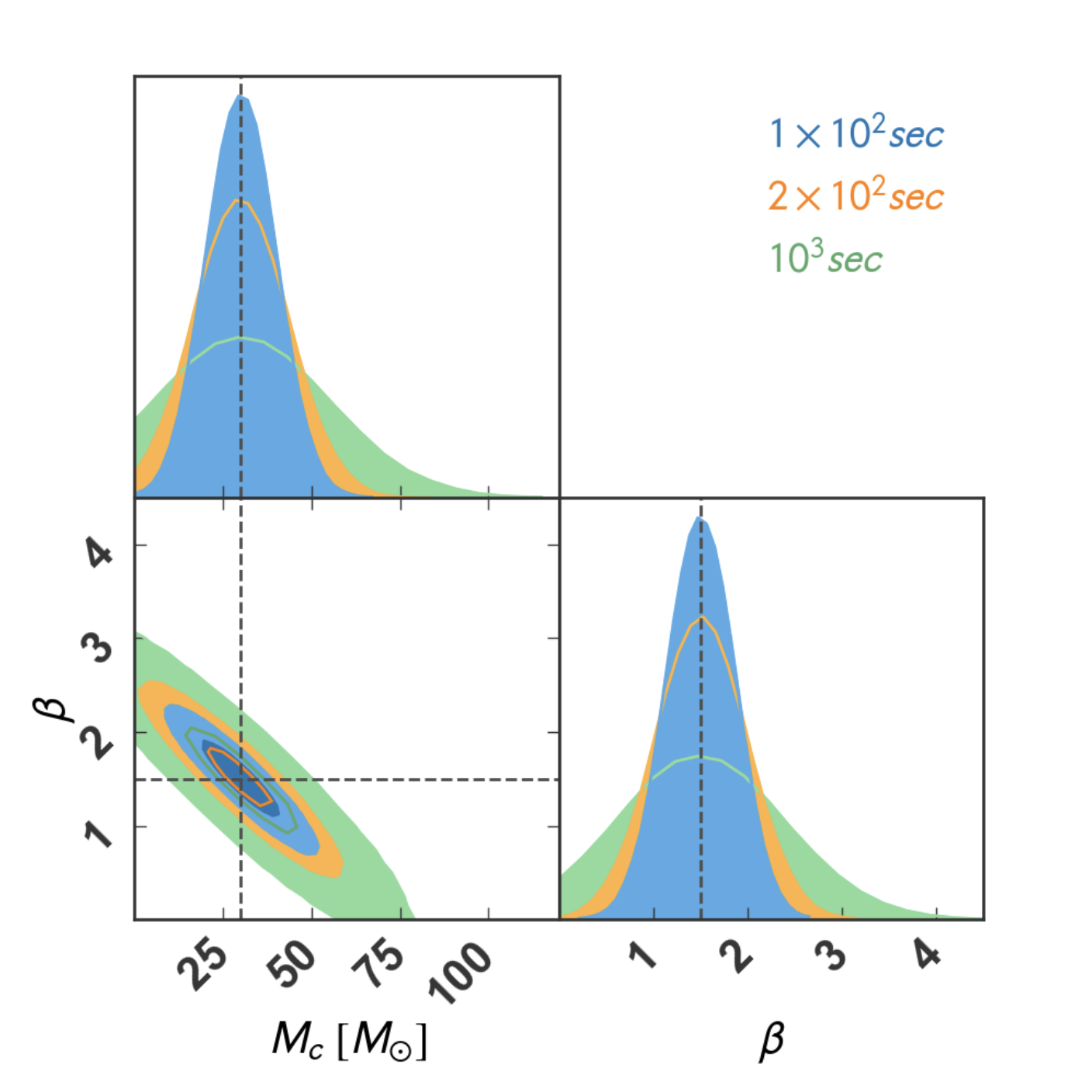}
    \caption{Corner plot showing how the measurement changes with the short-time bin ($\Delta T$) for PBHs with $M_c$= 30 $M_\odot$ and $\beta$= 1.5.}
    \label{gtcMc30-dt}
\end{figure}

The Figure of Merit (FoM) can be defined as the square root of the determinant of the Fisher matrix. It is a measure of the information content and effectiveness of the parameter estimation. In Fig. \ref{FoM-M}, we show the ratio of the FoM for the power spectrum + covariance case to that for the power spectrum-only case as a function of $M_c$. The ratio peaks near $M_c$ = 30 $M\odot$ and decreases for both higher and lower $M_c$ values. This is because the signal length ($\tau$) of low-mass sources is longer than that of high-mass sources. The longer signals make the background signal more stationary, thereby reducing the magnitude of the covariance matrix. On the other hand, high-mass sources can be individually detected up to higher redshifts, and the contribution to SGWB from higher-mass BHs comes from higher redshifts. Since the signals at higher redshifts are fainter, the strength of the covariance matrix decreases for the higher-mass sources. Consequently, the FoM is lower for both too-small and too-large $M_c$ values. In Fig. \ref{FoM-T}, we demonstrate the same ratio as a function of short-time bin, $\Delta$T. The ratio exhibits a decreasing trend with increasing short-time bin. As the size of the short-time bin width increases, the magnitude of the spectral covariance decreases, and therefore, the spectral covariance becomes increasingly insignificant when compared to the noise power spectrum. Moreover, beyond a short-time bin width size of $10^{3}$ seconds, the ratio converges and approaches 1. This convergence signifies that the effect of spectral covariance becomes essentially negligible beyond this threshold. The change in the bounds on the parameters with the bin size ($\Delta$T) is illustrated in Fig. \ref{gtcMc30-dt}, where we demonstrate how constraints on $M_c$ and $\beta$ change with $\Delta$T. The constraint improves as $\Delta$T decreases, reaching a saturation point beyond a bin size of around 200 seconds. The signal becomes statistically stationary when averaged over a large time. However, by measuring the non-stationary signal over a time window of $\Delta T$ seconds, and combining the signal from a large observational period, $T_{\rm obs}$, we can obtain a more accurate estimation of the parameters. This is equivalent to the spatial averaging of the fluctuations in the cosmic microwave temperature background. Cosmological fluctuations average to zero on average over a large spatial region leading to only a monopole signal. However, the fluctuations at small scales are prominent which can be measured from the variance of the fluctuation \citep{tegmark1997karhunen}.

\section{Conclusion}\label{conc}

In this paper, we developed a simulation suite of the SGWB signal and demonstrated how it can result in a non-stationary SGWB signal, depending on the properties of high-redshift BHs. This can lead to a non-zero correlation between different frequencies in the signal covariance matrix. We have shown that the SGWB will have both Gaussian and non-Gaussian signatures which are impacted differently by the merger rate and mass distribution. The study of non-Gaussian and non-stationary SGWB represents a new and promising approach for probing the high-redshift properties of BBH systems.

The inclusion of the spectral correlation in SGWB in our likelihood function can significantly improve our parameter estimation. This is because GW sources with different masses and merger rates lead to different structures in the covariance matrix of the SGWB signal. A larger merger rate will make the distribution more Gaussian-like, while the mass distribution will affect both the variance and skewness of the distribution. This is a key signature in distinguishing different mass distribution and merger rate models. We found that the technique can improve the measurement by up to 90\% with the A+ sensitivity of the detectors. In our analysis, we assumed that the noise at two different frequencies is uncorrelated, which is a good approximation for noise with a sufficiently long time scale over which it is stationary. It is worth noting that this analysis is moderately computationally expensive. The simulation of the $\Omega_{gw}(f)$ signal due to ABHs for one year of observation time with a short-time bin of $\Delta T= 200$ seconds requires approximately 90 core hours. Making the simulation computationally affordable for practical application to data within a Bayesian framework is going to be challenging.

Using the simulation-based approach to infer the signal covariance matrix makes it accessible for the first time to utilize the additional information present in the SGWB. Although computationally expensive, incorporating this approach into the estimation of the SGWB signal from data will be highly advantageous. We have demonstrated this by applying the technique to a physical model to assess its effectiveness in understanding the underlying population. In the future, we plan to apply this technique to the LVK data from the upcoming observations to explore the stellar properties of compact objects at high redshift and also investigate its performance in the presence of glitches 

Our method is a versatile technique that can be applied to a variety of different situations. Observations with future advanced gravitational-wave detectors will greatly enhance our ability to study non-stationary and non-Gaussian backgrounds and extract new insights into the BH and stellar properties at high redshifts. We will implement this theoretical framework in the future to explore the feasibility of understanding the BH population and its evolution with redshift for supermassive and intermediate-mass BHs from Pulsar Timing Array \citep{1990ApJ...361..300F, 2013CQGra..30v4008M, 2018JApA...39...51J,2013PASA...30...17M}, space-based GW detector LISA \citep{amaro2017laser,hughes2006brief}, space-based/moon-based GW detector in Deci-Hz frequency range \citep{Sato:2017dkf, Grimm:2020ivq} and low-mass BHs from Einstein Telescope \citep{punturo2010einstein,sathyaprakash2012scientific} and Cosmic Explorer \citep{hall2022cosmic,hall2021gravitational}. For some GW sources, LISA, Einstein Telescope, and Cosmic Explorer will achieve exceptional multi-band synergy and will be able to see deep up to a redshift of approximately z $\approx$ 50, and over a wide range of masses. The exploration of the time-dependent aspect of SGWB in the multi-band signal can bring a new window to understand the properties of the compact objects from high redshift which are vastly unexplored from electromagnetic observations. This technique can also be useful for distinguishing between astrophysical and cosmological SGWB signals based on the structure of the covariance matrix.

\section*{Acknowledgments}
The authors are thankful to Shivaraj Kandhasamy for reviewing the manuscript and providing useful comments.  This work is a part of the $\langle \texttt{data|theory}\rangle$ \texttt{Universe-Lab} which is supported by the TIFR and the Department of Atomic Energy, Government of India. 
 The authors would like to thank the  LIGO/Virgo scientific collaboration for providing the noise curves and the
computer center HPC facility at TIFR for providing computing resources. LIGO is funded by the U.S. National Science Foundation. Virgo is funded by the French Centre National de Recherche Scientifique (CNRS), the Italian Istituto Nazionale della Fisica Nucleare (INFN), and the Dutch Nikhef, with contributions by Polish and Hungarian institutes. This material is based upon work supported by NSF’s LIGO Laboratory which is a major facility fully funded by the National Science Foundation. The authors would also like to acknowledge the use of the following Python packages in this work: numpy \citep{van2011numpy}, scipy \citep{jones2001scipy}, matplotlib \citep{hunter2007matplotlib}, astropy \citep{robitaille2013astropy,price2018astropy}, pygtc \citep{bocquet2019pygtc}, and ray \citep{moritz2018ray}.
 
 \section*{Data Availability}
The data underlying this article will be shared at the request to the corresponding author.

\bibliographystyle{mnras}
\bibliography{main_paper}
\label{lastpage}
\end{document}